\documentclass[11pt]{article}
\usepackage[sc]{mathpazo}
\usepackage{fullpage}
\usepackage[authoryear,sectionbib,sort]{natbib}
\usepackage{graphicx}
\usepackage{adjustbox}
\usepackage{multirow}
\usepackage{amsmath}
\usepackage[utf8]{inputenc}
\usepackage{lineno}
\usepackage{titlesec}
\usepackage{graphicx}
\usepackage{subcaption}
\usepackage{booktabs,multirow}
\usepackage{here}
\usepackage{url}
\usepackage{algorithmic}
\usepackage{algorithm}

\newcommand{\mat}[1]{\mbox{\boldmath{$#1$}}} 

\titleformat{\section}[block]{\Large\bfseries\filcenter}{\thesection}{1em}{}
\titleformat{\subsection}[block]{\Large\itshape\filcenter}{\thesubsection}{1em}{}
\titleformat{\subsubsection}[block]{\large\itshape}{\thesubsubsection}{1em}{}
\titleformat{\paragraph}[runin]{\itshape}{\theparagraph}{1em}{}[. ]

\title{Asset price movement prediction using empirical mode decomposition and Gaussian mixture models}

\author{Gabriel R. Palma$^{1, 2,\ast}$ \and 
Mariusz Skoczeń$^{4}$ \and
Phil Maguire$^{3, 4}$}

\date{}

\begin{document}

\maketitle

\noindent{} 1. Hamilton Institute, Maynooth University, Maynooth, Ireland;

\noindent{} 2. Department of Mathematics and Statistics, Maynooth University, Maynooth, Ireland;

\noindent{} 3. Department of Computer Science, Maynooth University, Maynooth, Ireland;

\noindent{} 4. DLT Capital, Maynooth, Ireland

\noindent{} $\ast$ Corresponding author; e-mail: gabriel.palma.2022@mumail.ie

\bigskip

\bigskip


\bigskip

\textit{Manuscript type}: Research paper.

\bigskip

\noindent{\footnotesize Prepared using the suggested \LaTeX{} template for \textit{Am.\ Nat.}}

\newpage{}

\section*{Abstract}
We investigated the use of Empirical Mode Decomposition (EMD) combined with Gaussian Mixture Models (GMM), feature engineering and machine learning algorithms to optimize trading decisions. We used five, two, and one year samples of hourly candle data for GameStop, Tesla, and XRP (Ripple) markets respectively. Applying a 15 hour rolling window for each market, we collected several features based on a linear model and other classical features to predict the next hour's movement. Subsequently, a GMM filtering approach was used to identify clusters among these markets. For each cluster, we applied the EMD algorithm to extract high, medium, low and trend components from each feature collected. A simple thresholding algorithm was applied to classify market movements based on the percentage change in each market's close price. We then evaluated the performance of various machine learning models, including Random Forests (RF) and XGBoost, in classifying market movements. A naive random selection of trading decisions was used as a benchmark, which assumed equal probabilities for each outcome, and a temporal cross-validation approach was used to test models on $40\%$, $30\%$, and $20\%$ of the dataset. Our results indicate that transforming selected features using EMD improves performance, particularly for ensemble learning algorithms like Random Forest and XGBoost, as measured by accumulated profit. Finally, GMM filtering expanded the range of learning algorithm and data source combinations that outperformed the top percentile of the random baseline.

\newpage{}

\section{Introduction}
Predicting price movements in asset markets is challenging due to the inherent complexity and non-stationarity of financial time series \citep{zhang2009behavior, musaev2023genesis, hung2024ai, Deqin2014PredictionFN, purwantara2024deep, mutinda2025stock, chen2025accurate}. These challenges are amplified in diverse contexts, ranging from well-established equity markets to emergent and volatile domains such as cryptocurrencies, including well-known coins, such as Ripple (XRP) \citep{subburayan2024transforming, li2024stock, Tang2022FinancialTS, mutinda2025stock}. The nature of this problem motivates the creation of new quantitative approaches to predict market movements, including machine learning, statistics and classical algorithms ~\citep{sattar2025novel, rahman2025assessing}. 

Several researchers have invested in developing new features to aid market movement prediction~\citep{palma2024combining, wang2025multimodal, giri2025systematic}. One example is using classical candle patterns and time series features, such as moving averages and other metrics, as explanatory variables to develop quantitative methods~\citep{parente2024, praveen2025financial, narayana2025ensemble}. After extracting these features, decomposition techniques can be used to transform the feature space for further exploration using learning algorithms~\citep{bahri2022time}. This approach allows researchers to explore the feature space and disentangle structured signals from noise \citep{hung2024ai, Deqin2014PredictionFN, wang2023hybrid}.

A classical way of decomposing time series is using the Empirical Mode Decomposition (EMD)~\citep{Huang1998}. Several researchers have used this method when analysing non-linear and non-stationary time series \citep{rilling2007one, guhathakurta2008empirical, xuan2010empirical, Deqin2014PredictionFN, quinn2021emd, bahri2022time}. Overall, this method decomposes a time series into intrinsic mode functions (IMFs), which can be used for further analysis \citep{Deqin2014PredictionFN, quinn2021emd, xuan2010empirical}. This decomposition facilitates the separation of deterministic components, such as market trends, from stochastic elements, like high-frequency fluctuations \citep{Deqin2014PredictionFN, wang2023hybrid, riemenschneider2005b}. Several methods have been proposed utilising EMD and its extensions, such as Ensemble Empirical Mode Decomposition (EEMD) with different machine learning methods in various research areas, such as Long Short Term Memory (LSTM), Support Vector Machines, and others~\citep{Tang2022FinancialTS, mutinda2025stock, guo2025ensemble, nazari2025groundwater, zhu2025role}. The combination of EMD with supervised learning methods has shown positive results and motivated additional exploration of the potential of this technique \citep{wang2023hybrid, zhu2025role, subasi2025electroencephalography, someetheram2025hybrid}.

In the same context of feature engineering, Gaussian Mixture Models (GMM) have been applied in finance for clustering and identifying latent structures within markets, making them particularly suited for capturing the nuanced behaviours of financial time series \citep{zhang2020time, palma2024combining, saadaoui2025finite}. GMM has, in addition, been combined with machine learning techniques as a pre-processing tool used before classification tasks in several research areas, showing promising results in exploring the feature space and yielding higher learning algorithm performances~\citep{fan2022preprocessing, rajaguru2022gaussian, fan2023weighted}. In this context, GMM can be used to look for profitable trading opportunities within different markets~\citep{palma2024combining}.

Considering the promising results using both methodologies separately, its integration presents an opportunity for creating a unified approach combining machine learning, feature engineering, model-based clustering and time series decomposition. This study proposes combining EMD and GMM in a unified approach to predict market movements by leveraging features extracted from deterministic and different frequency levels of financial time series input signals. We apply this approach to the XRP, Tesla, and GameStop markets. The remainder of this paper is organised as follows: Section 2 details the data acquisition process, feature engineering methodology, time series decomposition, clustering approach using GMM, and the learning algorithms used in this paper. Section 3 presents the experimental results and discusses their implications. Finally, Section 4 presents the conclusions of this paper, introducing key findings and outlines directions for future research.

\section{Methods}
We collected approximately five, two and one year worth of hour-level data for the GameStop, Tesla, and XRP markets respectively. For each market, the closing price at hour $t$, denoted by $y_{\text{close}}(t)$, is used to compute the logarithmic return as follows:
$$\omega(t) = \log\frac{y_{\text{close}}(t+1)}{y_{\text{close}}(t)},$$
where $\omega(t)$ represents the percentage change in price \citep{parente2024, palma2024combining}. A threshold algorithm is then implemented to label the trading decisions. Thus, a selling decision is assigned if $\omega(t)$ is smaller than the lower quantile. We classify a buying decision if $\omega(t)$ exceeds the upper quantile. Otherwise, a holding decision is chosen \citep{palma2024combining}. We defined trading decisions for all markets by employing a simple threshold algorithm based on the $3.5\%$ quantile for our analysis, yielding balance classes among the time series~\citep{palma2024combining}.

After labelling each market based on the described algorithm, we used a sliding window of $15$ hours to extract features for training the learning algorithms to predict an hour ahead \citep{Buczynski2021, palma2024combining}. For each window and market, we obtained the set of features $X_i = x_i(t), x_i(t+1), \ldots, x_i(T_m)$, where $i = 1, \ldots, 8$ and represent, respectively, the buy proportion, sell proportion, close price, linear model intercept, linear model slope, peaks average curvature, peaks average magnitude, and estimated the percentage change for each market, $m$, with $T_m$ observations following the methods proposed by~\cite{palma2024combining}.  

We used a Gaussian Mixture Model (GMM) to assign a cluster to each set of features at a given time collected for each market separately. Let the GMM's probability density be defined as:
$$f(X_i, \Psi) = \sum_{g=1}^{G} \pi_g \phi(X_i; \mat{\mu}_g, \mat{\Sigma}_g),$$
where $G$ is the number the number of mixture components, $\phi(\cdot)$ is the multivariate Gaussian probability density function, $\mat{\mu}_g$ is the vector of means and $\mat{\Sigma}_g$ is the variance-covariance matrix. To avoid clusters with few observations, we tested GMMs with $G = 1, \ldots, 4$ and used the Bayesian Information Criteria (BIC) to select $G$ for a given market. By clustering the markets based on the selected features, $X_i$, we obtain groups with similar designed features, thus allowing further exploration with the decomposition of the set of features and applying machine learning algorithms in subsequent stages. We also evaluate the performance of learning algorithms without this step to further explore the effect of using the GMM-filtering step in our approach.

\begin{figure}
    \centering
    \includegraphics[width=0.8\linewidth]{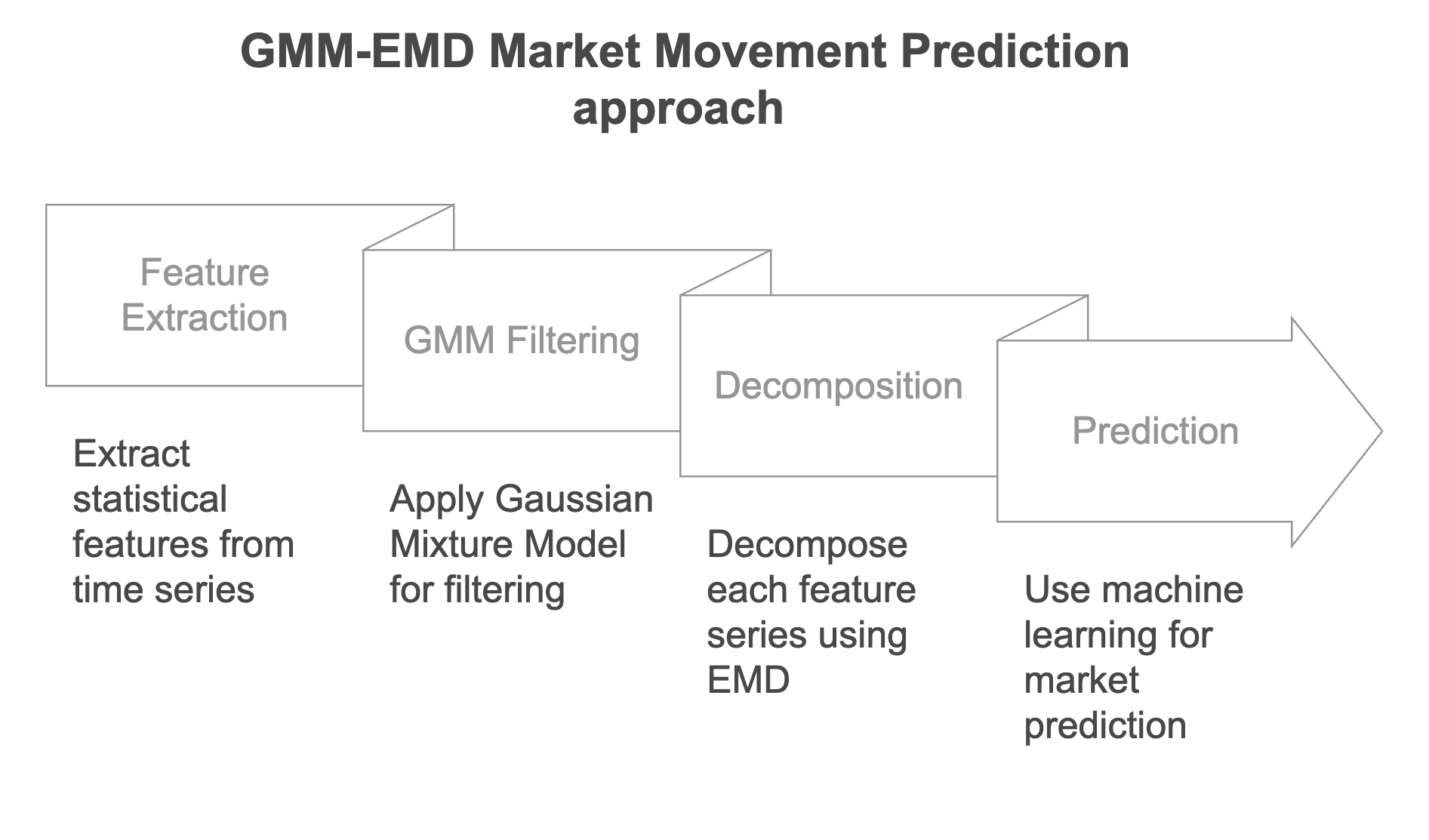}
    \caption{A diagram illustrating the complete proposed approach to predict market movements based on the features extracted from the selected series.}
    \label{Diagram}
\end{figure}

Subsequently, for each feature set $X_i$ filtered or not, we applied the Empirical Mode Decomposition (EMD) to separate these signals into \textit{Intrinsic Mode Functions} (IMFs) using a sifting process \citep{Huang1998, rilling2007one, bahri2022time}. An IMF must satisfy: (1) the number of extrema and zero-crossings differs by at most one, and (2) the mean of upper and lower envelopes is zero~\citep{Huang1998}. The algorithm iteratively extracts IMFs by:

\begin{equation}
X_i = \sum_{j=1}^J C_{i, j} + R_{i, J}
\end{equation}where $C_{i, j}$ is an IMF with observations $c_{i, j}(t)$ and $j = \{1, \ldots, J\}$ is the index of the IMFs ordered from highest to lowest frequency. $R_{i, J}$ is the time series of residuals with observations $r_{i, j}(t)$. The residual is obtained by a sifting process briefly described in Algorithm~1. 

\begin{algorithm}
\caption{Empirical Mode Decomposition (EMD) pseudo algorithm applied to $X_1$.}
\begin{algorithmic}[1]
\STATE \textbf{Let:} $X_1$ be series of the buy proportions and $H_{j, k}$ be the series containing observations, $h_{j, k}(t)$, of the sifting procedure used to compute $j^\text{th}$ IMF;
\STATE \textbf{Set:} $k = 1$ and $j = 1$;
\STATE Construct upper $e_{\text{up}}$ and lower $e_{\text{low}}$ envelopes via cubic spline, and mean envelop $m_{j, k} = \frac{1}{2}(e_{\text{up}} + e_{\text{low}})$ based on the local maxima/minima of $X_1$;
\STATE Compute observations of the sifting procedure $h_{j, k}(t) = x_1(t) - m_{j, k}$;
\WHILE{$H_{j, k}$ has $\geq 2$ extrema}    
    \REPEAT        
        \STATE Identify local maxima/minima of $H_{j, k}$
        \STATE Construct upper $e_{\text{up}}$ and lower $e_{\text{low}}$ envelopes via cubic spline
        \STATE Compute mean envelope: $m_{j, k} = \frac{1}{2}(e_{\text{up}} + e_{\text{low}})$
        \STATE $h_{j, k+1}(t) = h_{j, k}(t) - m_{j, k}$        
        \STATE $k = k + 1$
    \UNTIL{Stopping criterion: $0.2 < \displaystyle\sum_{t}^{T} \frac{|h_{j, k-1}(t)-h_{j, k}(t)|^2}{h_{j, k-1}^2(t)} < 0.3$} 
    \STATE $c_{1, j}(t) = h_{j, k}(t)$ \COMMENT{Store $j^{\text{th}}$ IMF}
    \IF{$j = 1$}
        \STATE $r_{1, j}(t) = x_{1}(t) - c_{1, j}(t)$
    \ELSIF{$j > 1$}
        \STATE $r_{1, j}(t) = r_{1, j - 1}(t) - c_{1, j}(t)$    
    \ENDIF    
    \STATE $j = j + 1$
\ENDWHILE
\STATE \textbf{Return:} $J$ IMFs $\{C_{1, 1}, C_{1, 2}, \ldots, C_{1, J}\}$ and residuals $\{ R_{1, J} = r_{1, j}(t), r_{1, j}(t + 1), \ldots, r_{1, j}(T))\}$
\end{algorithmic}
\end{algorithm}

Following the EMD decomposition process, we define distinct components representing trends and varying levels of stochasticity for analysing market dynamics. For each feature $X_i$ with $J$ total IMFs extracted through the EMD process, we establish three adaptive cutoff indices: $r_1 = \max(1, J-6)$, $r_2 = \max(1, J-4)$, and $r_3 = \max(1, J-2)$. The trend (deterministic component) is defined as $\sum_{j=r_3+1}^{J} C_{i,j}$, representing the low-frequency underlying pattern of the time series. We characterise stochasticity in three distinct levels: high stochasticity as $\sum_{j=1}^{r_1} C_{i,j}$, medium stochasticity as $\sum_{j=1}^{r_2} C_{i,j}$, and low stochasticity as $\sum_{j=1}^{r_3} C_{i,j}$. 

Once the data are clustered and we employ the EMD for each feature set, supervised machine-learning models are trained within each cluster to predict trading decisions. Also, we include the non-clustered dataset to evaluate the effect on the algorithms' performance compared to the random algorithm. This paper included the Random Forests, K-Nearest Neighbours (KNN), Support Vector Machines (SVM) with the polynomial kernel, Deep Neural Networks and XGBoost algorithms to predict market movements. A simple random algorithm was used as a baseline to compare the performances across different markets and clusters of those markets \citep{parente2024}. The performance is evaluated using a financial performance metric, the Accumulated Percentage Change ($APC$), defined as:

$$APC = \sum_{t=1}^{T-1} g\left(\omega(t+1), d(t)\right),$$

where $d(t)$ is the predicted trading decision at time \(t\) and $g(\cdot)$ is a profit function specific to each action \citep{palma2024combining}. A temporal cross-validation strategy is adopted by partitioning the data into various training and testing sets to ensure robust performance across different market regimes. We presented the results for testing sets of $20\%$, $30\%$, and $40\%$ of the market or GMM-filtered market. The Python and R programming languages were used to implement the methods proposed in this paper. To allow full reproducibility of the findings, we have made the code available at \url{https://github.com/GabrielRPalma/GMMEMDForecast}. We also developed a new Python package entitled \texttt{fmfeatures} that includes the main functions used in this paper and is available at \url{https://pypi.org/project/fmfeatures/}. This package will continue to be updated as a tool for helping Python developers implement the techniques presented in this paper.

\section{Results and discussion}
\begin{figure}
    \centering
    \includegraphics[width=1\linewidth]{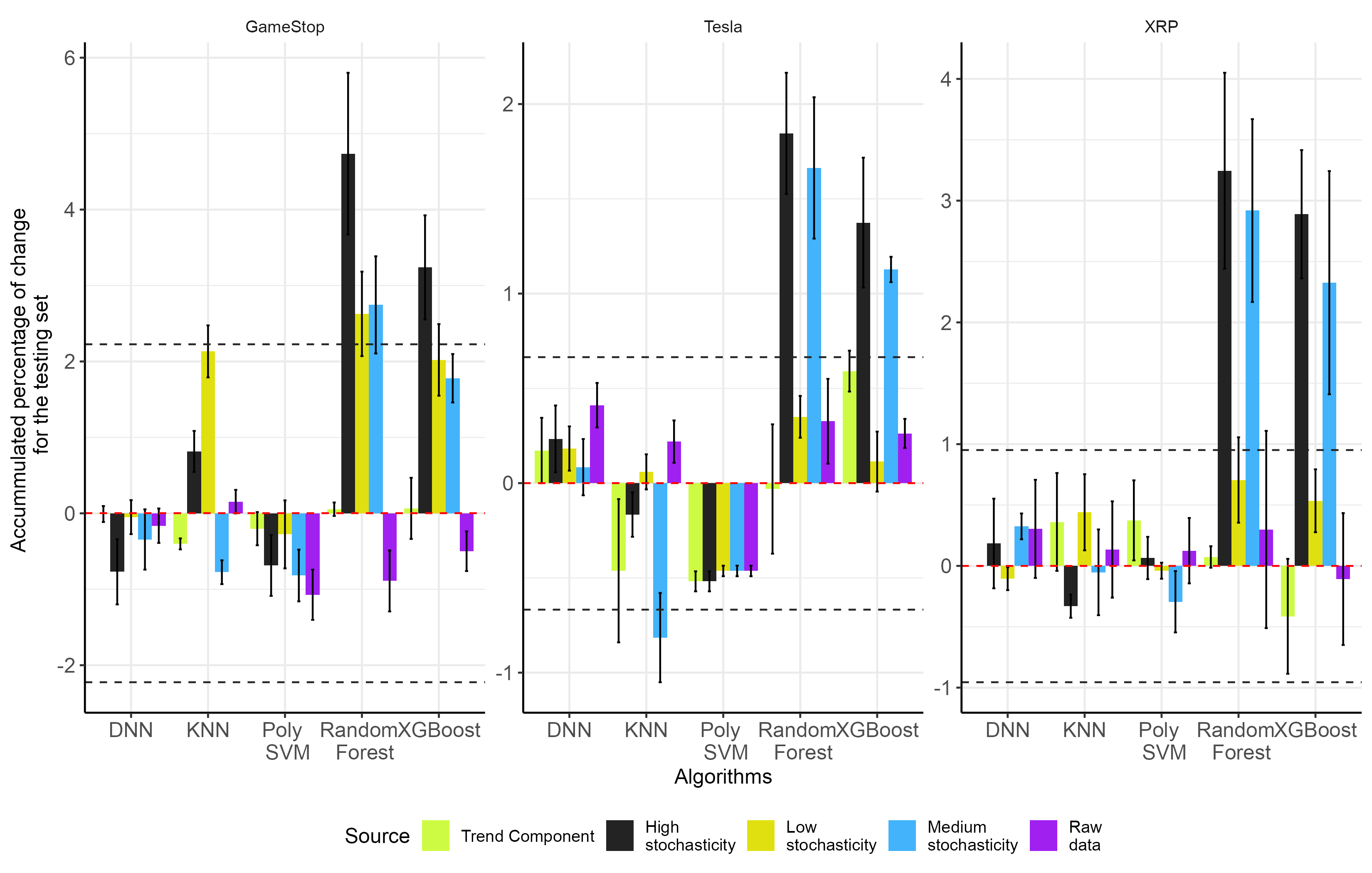}
    \caption{Accumulated profit ($APC$) obtained by the learning algorithms per studied market, when using the EMD components and the raw features. The red and black dashed lines represent the average, the $2.5\%$ and $97.5\%$ percentiles of the performance metric using the random algorithm.}
    \label{EMDPerformanceAllMarkets}
\end{figure}
Figure~\ref{EMDPerformanceAllMarkets} shows the performance of the machine learning algorithms obtained for the classification of trading decisions for all markets based on the raw series of features, the trend, high stochasticity, low stochasticity and medium stochasticity components, as defined based on the EMD. Considering all sources and algorithms, the average $APC$ obtained for the XRP market was $0.56$, with a standard deviation of $(1.22)$. For the GameStop market, the average $APC$ was $0.54$ ($1.64$), while for the Tesla market the average $APC$ was lower, at $0.20$ ($0.73$), indicating lower returns than for GameStop and XRP. Considering all the markets and sources, the average $APC$ obtained for the learning algorithms Deep Neural Networks, K-Nearest Neighbours, Polynomial Support Vector Machines, Random Forests, and XGBoost were, respectively, $0.03$ ($0.45$), $0.09$ ($0.80$), $-0.35$  ($0.51$), $1.38$  ($1.73$), and $1.02$ ($1.33$), indicating that the ensemble-based algorithms, Random Forests and XGBoost produced a higher performance than the other algorithms. Considering all algorithms and markets, the average $APC$ for the high, medium, and low stochasticity, trend and raw time series were, respectively, $1.08$ ($1.82$), $0.63$ ($1.44$), $0.55$ ($1.01$), $-0.023$ ($0.49$), and $-0.064$ ($0.67$). 

Figure~\ref{EMDPerformanceAllMarkets} shows that for GameStop, Tesla, and XRP, four combinations of algorithms and sources produced higher performance than the upper percentile of the random baseline. The ensemble-based algorithms, Random Forests and XGBoost, combined with the empirical mode decomposition's high and medium stochasticity components, yielded higher $APC$ values than the $97.5\%$ percentile of the random baseline. This finding highlights the impact of the application of the empirical mode decomposition method on the learning algorithm's performance, agreeing with previous findings on the use of EMD and EMD extension methods, such as Ensemble Empirical Mode Decomposition (EEMD), applied to financial time series ~\citep{xu2023emd, li2024stock, yu2024carbon, suo2024driver}.

\begin{figure}
    \centering
    \includegraphics[width=1\linewidth]{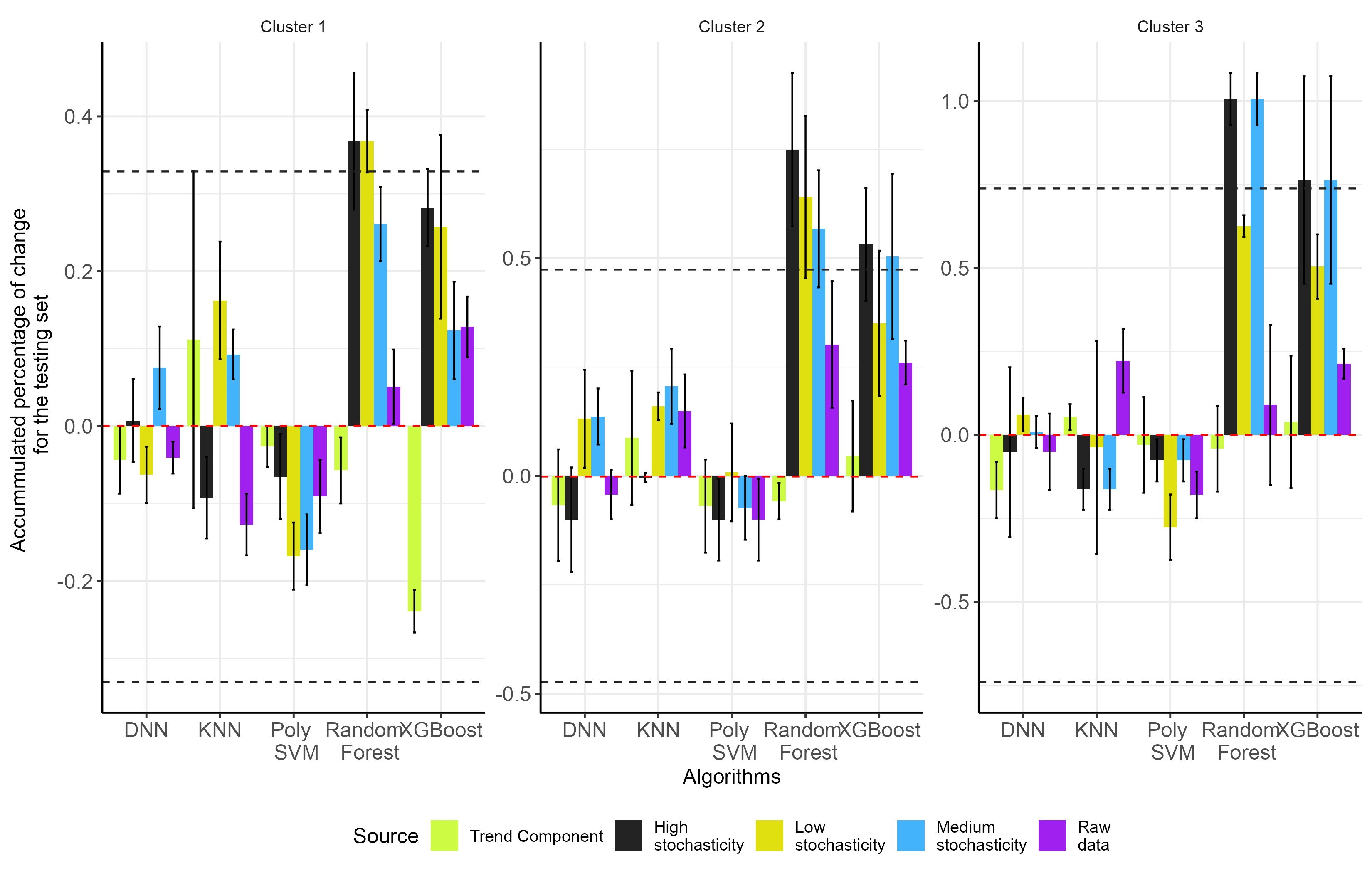}
    \caption{Accumulated profit ($APC$) obtained by the learning algorithms per cluster presented in the XRP market, when using the EMD components and the raw features. The red and black dashed lines represent the average, the $2.5\%$ and $97.5\%$ percentiles of the performance metric using the random basline.}
    \label{XRPPerformance}
\end{figure}

Figure~\ref{XRPPerformance} presents the performance of the learning algorithms after the GMM-filtering procedure, including the effect of the empirical mode decomposition on the extracted features of the XRP market. Overall, considering all clusters and sources, the average $APC$ obtained for the Deep Neural Networks, K-Nearest Neighbours, Polynomial Support Vector Machines, Random Forests, and XGBoost were, respectively, $-0.01$ ($0.16$), $0.04$ ($0.22$), $-0.10$ ($0.14$), $0.39$ ($0.39$), and $0.30$ ($0.35$) highlighting that the ensemble-based algorithms performed better in this market. Considering all algorithms and clusters, the average $APC$ for the high, medium, and low stochasticity, trend and raw time series were, respectively, $0.20$ ($0.42$), $0.22$ ($0.38$), $0.18$ ($0.32$), $-0.03$ ($0.19$), and $0.05$ ($0.20$).

Considering all learning algorithms and sources, the average $APC$ obtained for the first, second and third clusters were, respectively, $0.04$ ($0.19$), $0.17$ ($0.30$), $0.16$ ($0.43$). Moreover, the second cluster contains five combinations of algorithms and sources that produced higher performance than the upper percentile of the random baseline. The ensemble-based algorithms combined with the high, medium and low stochasticity components increased the possible approaches to obtain higher $APC$ compared to the baseline. The XRP performances presented in Figure~\ref{EMDPerformanceAllMarkets} highlight four combinations that outperformed the baseline, indicating promising results for combining the GMM and EMD methods into the proposed framework.

\begin{figure}
    \centering
    \includegraphics[width=1\linewidth]{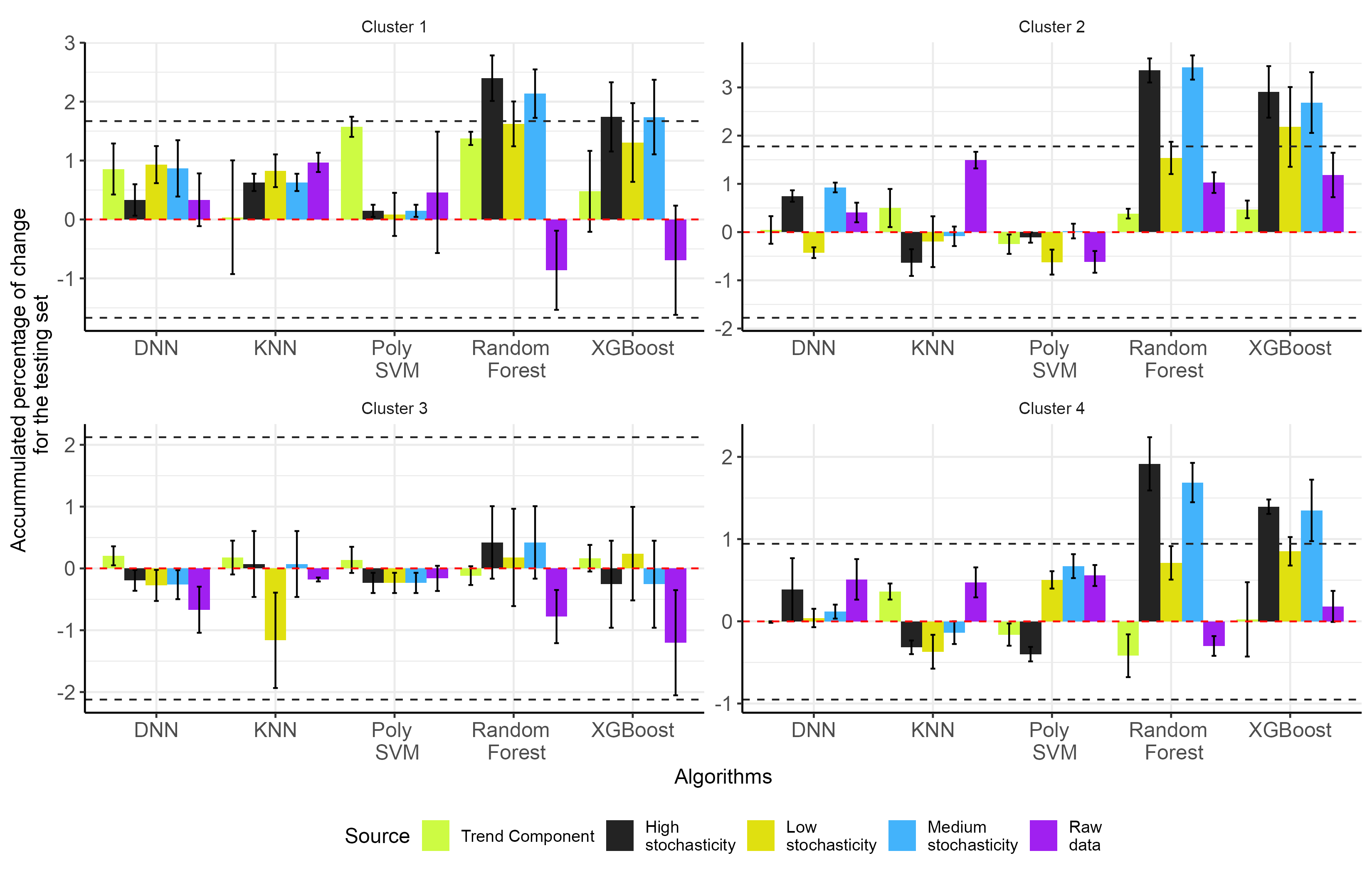}
    \caption{Accumulated profit ($APC$) obtained by the learning algorithms per cluster presented in the GameStop market, when using the EMD components and the raw features. The red and black dashed lines represent the average, the $2.5\%$ and $97.5\%$ percentiles of the performance metric using the random algorithm.}
    \label{GameStopPerformance}
\end{figure}

Figure~\ref{GameStopPerformance} shows the performance of the learning algorithms after the GMM-filtering procedure, including the effect of the empirical mode decomposition on the extracted features of the GameStop market. Overall, considering all clusters and sources the average $APC$ obtained for the Deep Neural Networks, K-Nearest Neighbours, Polynomial Support Vector Machines, Random Forests, and XGBoost were, respectively, $0.24$ ($0.61$), $0.16$ ($0.81$), $0.064$ ($0.65$), $1.00$ ($1.36$), and $0.82$ ($1.37$). These results also emphasise the higher performance obtained by the ensemble-based algorithms in this market. Considering all algorithms and clusters, the average $APC$ for the high, medium, and low stochasticity, trend and raw time series were, respectively, $0.71$ ($1.26$), $0.79$ ($1.17$), $0.39$ ($1.06$), $0.29$ ($0.70$), and $0.11$ ($0.99$).

Considering all learning algorithms and sources, the average $APC$ obtained for the first, second, third and fourth clusters were, respectively, $0.80$ ($1.08$), $0.81$ ($1.32$), $-0.16$ ($0.79$), and $0.38$ ($0.71$). The first and second clusters contain four and five combinations of algorithms and sources that produced higher performance than the upper percentile of the random baseline. In addition, for cluster one, more learning algorithms obtained positive $APC$ values despite being less than the upper bound of the random baseline. The same is noticeable for the second cluster. However, the polynomial Support Vector Machines had similar performances compared to the non-filtered approach for the GameStop market presented in Figure~\ref{EMDPerformanceAllMarkets}.

\begin{figure}
    \centering
    \includegraphics[width=1\linewidth]{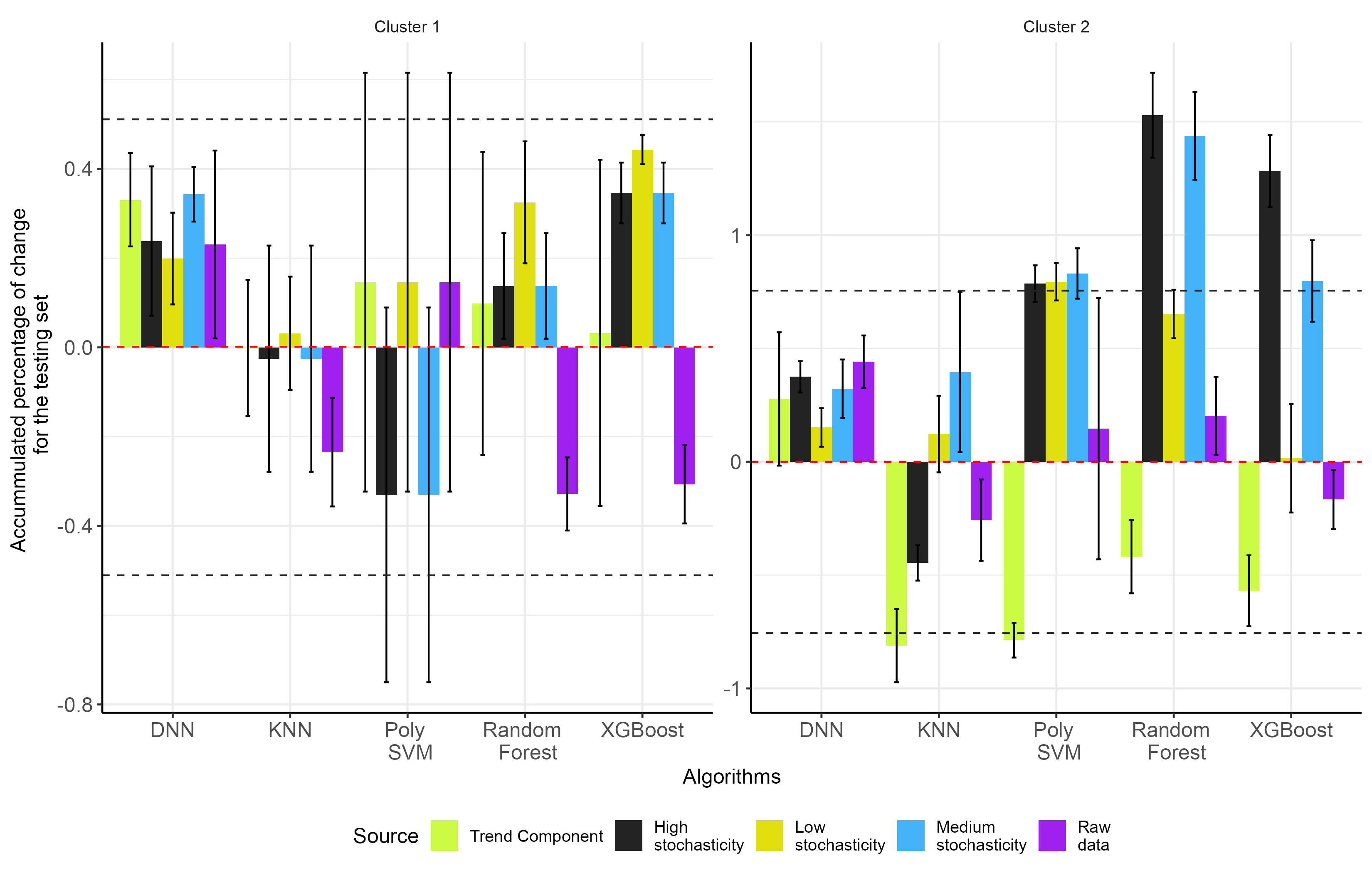}
    \caption{Accumulated profit ($APC$) obtained by the learning algorithms per cluster presented in the Tesla market, when using the EMD components and the raw features. The red and black dashed lines represent the average, the $2.5\%$ and $97.5\%$ percentiles of the performance metric using the random algorithm.}
    \label{TeslaPerformance}
\end{figure}

Figure~\ref{TeslaPerformance} shows the performance of the learning algorithms after the GMM-filtering procedure, including the effect of the empirical mode decomposition on the extracted features of the Tesla market. Overall, considering all clusters and sources the average $APC$ obtained for the Deep Neural Networks, K-Nearest Neighbours, Polynomial Support Vector Machines, Random Forests, and XGBoost were, respectively, $0.29$ ($0.23$), $-0.12$ ($0.43$), $0.15$ ($0.75$), $0.38$ ($0.68$), and $0.22$ ($0.59$). The Tesla market did not distinguish clearly among learning algorithms from the previous markets. Considering all algorithms and clusters, the average $APC$ for the high, medium, and low stochasticity, trend and raw time series were, respectively, $0.39$ ($0.68$), $0.43$ ($0.57$), $0.29$ ($0.38$), $-0.17$ ($0.56$), and $-0.01$ ($0.47$).

Considering all learning algorithms and sources presented in Figure~\ref{TeslaPerformance}, the average $APC$ obtained for the first and second clusters were, respectively, $0.08$ ($0.43$) and $0.28$ ($0.69$), indicating that the second cluster of the Tesla market provided more profitable opportunities. It contains seven combinations of algorithms and sources that produced higher performance than the upper percentile of the random baseline. Compared to the four combinations presented by the non-filtered tesla market in Figure~\ref{EMDPerformanceAllMarkets}, this is a promising result for further exploration of the GMM-filtering approach, and it agrees with other findings in the literature~\citep{fan2022preprocessing, palma2024combining}.

\section{Conclusion}
We explored using feature engineering, Gaussian mixture models, empirical mode decomposition, and classical machine learning algorithms to predict market movements. The performances obtained based on $APC$ showed the feasibility of the proposed approach and highlighted previous findings in the literature related to the use of EMD and EEMD on financial markets~\citep{Tang2022FinancialTS, xu2023emd, palma2024combining}. As this study serves as the basis for further implementations of the proposed approach, we explored the learning algorithms' performance on already clustered markets, and did not include the classification error for the GMM's classes. As the results indicate the potential of finding additional profits based on the studied clusters, we will include the GMM classification error as part of the approach in future work. In addition, this framework introduces the possibility of studying multiple markets using different signals owing to the use of empirical mode decomposition. 

Our results have shown that using the EMD increases accumulated profit, $APC$, solely for ensemble-based algorithms. The GMM allows further exploration of the feature space, producing an increased number of combinations of learning algorithms and sources that yielded higher performance than the upper percentile of the random baseline. It allowed the learning algorithms to find more profitable opportunities in the clustered markets for all three datasets. In future work we will continue exploring these techniques and adapting learning algorithms focused on data stream learning tasks,including the addition of portfolio optimisation among sets of assets.

\section{Acknowledgments}

This publication has resulted from research conducted with the financial support of DLT Capital and Taighde Éireann – Research Ireland under Grant 18/CRT/6049. The opinions, findings and conclusions or recommendations expressed in this material are those of the authors and do not necessarily reflect the views of the funding agencies.

\section{Declarations}
~~~~
\textbf{Ethical Approval} Not applicable.

\textbf{Competing interests} DLT Capital's interests did not influence the impartiality of this study, and we confirm that there is no conflict of interest.

\textbf{Funding} Taighde Éireann – Research Ireland under Grant 18/CRT/6049 and DLT Capital.

\bibliographystyle{apalike}
\bibliography{ref}

\begin{thebibliography}{}

\bibitem[Bahri and Vahidnia, 2022]{bahri2022time}
Bahri, M.~Z. and Vahidnia, S. (2022).
\newblock Time series forecasting using smoothing ensemble empirical mode decomposition and machine learning techniques.
\newblock In {\em 2022 International Conference on Electrical, Computer, Communications and Mechatronics Engineering (ICECCME)}, pages 1--6. IEEE.

\bibitem[Buczynski et~al., 2021]{Buczynski2021}
Buczynski, W., Cuzzolin, F., and Sahakian, B. (2021).
\newblock A review of machine learning experiments in equity investment decision-making: why most published research findings do not live up to their promise in real life.
\newblock {\em International Journal of Data Science and Analytics}, 11:221--242.

\bibitem[Chen et~al., 2025]{chen2025accurate}
Chen, X., Tang, G., Ren, Y., Lin, X., and Li, T. (2025).
\newblock Accurate and efficient stock market index prediction: an integrated approach based on vmd-snns.
\newblock {\em Journal of Applied Statistics}, 52(4):841--867.

\bibitem[De-qin, 2014]{Deqin2014PredictionFN}
De-qin, W. (2014).
\newblock Prediction for non-stationary non-linear time series based on empirical mode decomposition.
\newblock {\em Systems Engineering}.

\bibitem[Fan et~al., 2022]{fan2022preprocessing}
Fan, C., Zhang, N., Jiang, B., and Liu, W.~V. (2022).
\newblock Preprocessing large datasets using gaussian mixture modelling to improve prediction accuracy of truck productivity at mine sites.
\newblock {\em Archives of Mining Sciences}, pages 661--680.

\bibitem[Fan et~al., 2023]{fan2023weighted}
Fan, C., Zhang, N., Jiang, B., and Liu, W.~V. (2023).
\newblock Weighted ensembles of artificial neural networks based on gaussian mixture modeling for truck productivity prediction at open-pit mines.
\newblock {\em Mining, Metallurgy \& Exploration}, 40(2):583--598.

\bibitem[Giri et~al., 2025]{giri2025systematic}
Giri, S., Du, D., and Beruvides, M. (2025).
\newblock A systematic approach to predicting nft prices using time series forecasting and macroeconomic factors in digital assets.
\newblock {\em Cogent Economics \& Finance}, 13(1):2468387.

\bibitem[Guhathakurta et~al., 2008]{guhathakurta2008empirical}
Guhathakurta, K., Mukherjee, I., and Chowdhury, A.~R. (2008).
\newblock Empirical mode decomposition analysis of two different financial time series and their comparison.
\newblock {\em Chaos, Solitons \& Fractals}, 37(4):1214--1227.

\bibitem[Guo et~al., 2025]{guo2025ensemble}
Guo, Y., Si, J., Wang, Y., Hanif, F., Li, S., Wu, M., Xu, M., and Mi, J. (2025).
\newblock Ensemble-empirical-mode-decomposition (eemd) on swh prediction: The effect of decomposed imfs, continuous prediction duration, and data-driven models.
\newblock {\em Ocean Engineering}, 324:120755.

\bibitem[Huang et~al., 1998]{Huang1998}
Huang, N.~E., Shen, Z., Long, S.~R., Wu, M.~C., Shih, H.~H., Zheng, Q., Yen, N.-C., Tung, C.~C., and Liu, H.~H. (1998).
\newblock The empirical mode decomposition and the hilbert spectrum for nonlinear and non-stationary time series analysis.
\newblock {\em Proceedings of the Royal Society of London. Series A: mathematical, physical and engineering sciences}, 454(1971):903--995.

\bibitem[Hung et~al., 2024]{hung2024ai}
Hung, M.-C., Chen, A.-P., and Yu, W.-T. (2024).
\newblock Ai-driven intraday trading: Applying machine learning and market activity for enhanced decision support in financial markets.
\newblock {\em IEEE Access}, 12:12953--12962.

\bibitem[Li et~al., 2024]{li2024stock}
Li, S., Tang, G., Chen, X., and Lin, T. (2024).
\newblock Stock index forecasting using a novel integrated model based on ceemdan and tcn-gru-cbam.
\newblock {\em IEEE Access}.

\bibitem[Musaev et~al., 2023]{musaev2023genesis}
Musaev, A., Makshanov, A., and Grigoriev, D. (2023).
\newblock The genesis of uncertainty: Structural analysis of stochastic chaos in finance markets.
\newblock {\em Complexity}, 2023(1):1302220.

\bibitem[Mutinda and Geletu, 2025]{mutinda2025stock}
Mutinda, J.~K. and Geletu, A. (2025).
\newblock Stock market index prediction using ceemdan-lstm-bpnn-decomposition ensemble model.
\newblock {\em Journal of Applied Mathematics}, 2025(1):7706431.

\bibitem[Narayana et~al., 2025]{narayana2025ensemble}
Narayana, M.~L., Kartha, A.~J., Mandal, A.~K., Suresh, A., Jose, A.~C., et~al. (2025).
\newblock Ensemble time series models for stock price prediction and portfolio optimization with sentiment analysis.
\newblock {\em Journal of Intelligent Information Systems}, pages 1--25.

\bibitem[Nazari et~al., 2025]{nazari2025groundwater}
Nazari, A., Jamshidi, M., Roozbahani, A., and Golparvar, B. (2025).
\newblock Groundwater level forecasting using empirical mode decomposition and wavelet-based long short-term memory (lstm) neural networks.
\newblock {\em Groundwater for Sustainable Development}, 28:101397.

\bibitem[Palma et~al., 2024]{palma2024combining}
Palma, G.~R., Maguire, P., et~al. (2024).
\newblock Combining supervised and unsupervised learning methods to predict financial market movements.
\newblock {\em arXiv preprint arXiv:2409.03762}.

\bibitem[Parente et~al., 2024]{parente2024}
Parente, M., Rizzuti, L., and Trerotola, M. (2024).
\newblock A profitable trading algorithm for cryptocurrencies using a neural network model.
\newblock {\em Expert Systems with Applications}, 238:121806.

\bibitem[Praveen et~al., 2025]{praveen2025financial}
Praveen, M., Dekka, S., Sai, D.~M., Chennamsetty, D.~P., and Chinta, D.~P. (2025).
\newblock Financial time series forecasting: A comprehensive review of signal processing and optimization-driven intelligent models.
\newblock {\em Computational Economics}, pages 1--27.

\bibitem[Purwantara et~al., 2024]{purwantara2024deep}
Purwantara, I. M.~A., Setyanto, A., Utami, E., et~al. (2024).
\newblock Deep learning in financial markets: A systematic literature review of methods and future direction for price prediction.
\newblock In {\em 2024 6th International Conference on Cybernetics and Intelligent System (ICORIS)}, pages 01--06. IEEE.

\bibitem[Quinn et~al., 2021]{quinn2021emd}
Quinn, A.~J., Lopes-dos Santos, V., Dupret, D., Nobre, A.~C., and Woolrich, M.~W. (2021).
\newblock Emd: Empirical mode decomposition and hilbert-huang spectral analyses in python.
\newblock {\em Journal of open source software}, 6(59):2977.

\bibitem[Rahman et~al., 2025]{rahman2025assessing}
Rahman, M.~K., Dalim, H.~M., Reza, S.~A., Ahmed, A., Zeeshan, M. A.~F., Jui, A.~H., and Nayeem, M.~B. (2025).
\newblock Assessing the effectiveness of machine learning models in predicting stock price movements during energy crisis: Insights from shell's market dynamics.
\newblock {\em Journal of Business and Management Studies}, 7(1):44--61.

\bibitem[Rajaguru et~al., 2022]{rajaguru2022gaussian}
Rajaguru, H., SR, S.~C., and Chidambaram, S. (2022).
\newblock Gaussian mixture model based hybrid machine learning for lung cancer classification using symptoms.
\newblock In {\em 2022 Smart Technologies, Communication and Robotics (STCR)}, pages 1--4. IEEE.

\bibitem[Riemenschneider et~al., 2005]{riemenschneider2005b}
Riemenschneider, S., Liu, B., Xu, Y., and Huang, N.~E. (2005).
\newblock B-spline based empirical mode decomposition.
\newblock {\em The Hilbert-Huang transform and its applications. Singapur: World Scientific Publishing Co}, pages 27--56.

\bibitem[Rilling and Flandrin, 2007]{rilling2007one}
Rilling, G. and Flandrin, P. (2007).
\newblock One or two frequencies? the empirical mode decomposition answers.
\newblock {\em IEEE transactions on signal processing}, 56(1):85--95.

\bibitem[Sa{\^a}daoui and Rabbouch, 2025]{saadaoui2025finite}
Sa{\^a}daoui, F. and Rabbouch, H. (2025).
\newblock Finite mixture models for multiscale dynamics in chinese financial markets.
\newblock {\em Journal of the Operational Research Society}, 76(1):97--110.

\bibitem[Sattar et~al., 2025]{sattar2025novel}
Sattar, A., Sarwar, A., Gillani, S., Bukhari, M., Rho, S., and Faseeh, M. (2025).
\newblock A novel rms-driven deep reinforcement learning for optimized portfolio management in stock trading.
\newblock {\em IEEE Access}.

\bibitem[Someetheram et~al., 2025]{someetheram2025hybrid}
Someetheram, V., Marsani, M.~F., Kasihmuddin, M. S.~M., Jamaludin, S. Z.~M., Mansor, M.~A., and Zamri, N.~E. (2025).
\newblock Hybrid double ensemble empirical mode decomposition and k-nearest neighbors model with improved particle swarm optimization for water level forecasting.
\newblock {\em Alexandria Engineering Journal}, 115:423--433.

\bibitem[Subasi et~al., 2025]{subasi2025electroencephalography}
Subasi, A., Subasi, M.~E., and Qaisar, S.~M. (2025).
\newblock Electroencephalography-based emotion recognition with empirical mode decomposition and ensemble machine learning methods.
\newblock In {\em Artificial Intelligence Applications for Brain--Computer Interfaces}, pages 183--203. Elsevier.

\bibitem[Subburayan et~al., 2024]{subburayan2024transforming}
Subburayan, B., Sankarkumar, A.~V., Singh, R., and Mushi, H.~M. (2024).
\newblock Transforming of the financial landscape from 4.0 to 5.0: Exploring the integration of blockchain, and artificial intelligence.
\newblock {\em Applications of block chain technology and artificial intelligence}, pages 137--161.

\bibitem[Suo et~al., 2024]{suo2024driver}
Suo, R., Wang, Q., and Han, Q. (2024).
\newblock Driver analysis and integrated prediction of carbon emissions in china using machine learning models and empirical mode decomposition.
\newblock {\em Mathematics}, 12(14):2169.

\bibitem[Tang, 2022]{Tang2022FinancialTS}
Tang, K. (2022).
\newblock Financial time series prediction based on emd-svm.
\newblock {\em BCP Business \& Management}.

\bibitem[Wang et~al., 2023]{wang2023hybrid}
Wang, C., Yang, Y., Xu, L., and Wong, A. (2023).
\newblock A hybrid model of primary ensemble empirical mode decomposition and quantum neural network in financial time series prediction.
\newblock {\em Fluctuation and Noise Letters}, 22(04):2340006.

\bibitem[Wang et~al., 2025]{wang2025multimodal}
Wang, H., Xie, Z., Chiu, D.~K., and Ho, K.~K. (2025).
\newblock Multimodal market information fusion for stock price trend prediction in the pharmaceutical sector.
\newblock {\em Applied Intelligence}, 55(1):1--27.

\bibitem[Xu and He, 2023]{xu2023emd}
Xu, T. and He, X. (2023).
\newblock Emd-bilstm stock price trend forecasting model based on investor sentiment.
\newblock {\em Front. Comput. Intell. Syst}, 4(3):139--143.

\bibitem[Xuan et~al., 2010]{xuan2010empirical}
Xuan, Z., Xie, S., and Sun, Q. (2010).
\newblock The empirical mode decomposition process of non-stationary signals.
\newblock In {\em 2010 International Conference on Measuring Technology and Mechatronics Automation}, volume~3, pages 866--869. IEEE.

\bibitem[Yu et~al., 2024]{yu2024carbon}
Yu, Q., Abdul~Rahman, R., and Wu, Y. (2024).
\newblock Carbon price prediction in china based on ensemble empirical mode decomposition and machine learning algorithms.
\newblock {\em Environmental Science and Pollution Research}, pages 1--15.

\bibitem[Zhang et~al., 2009]{zhang2009behavior}
Zhang, J., Wang, J., and Shao, J. (2009).
\newblock The behavior of stock markets and the option pricing by the dynamic systems.
\newblock In {\em Proceedings of the 8th WSEAS international conference on Instrumentation, measurement, circuits and systems}, pages 200--205.

\bibitem[Zhang et~al., 2020]{zhang2020time}
Zhang, S., Tao, M., Niu, X.-F., and Huffer, F. (2020).
\newblock Time-varying gaussian-cauchy mixture models for financial risk management.
\newblock {\em arXiv preprint arXiv:2002.06102}.

\bibitem[Zhu, 2025]{zhu2025role}
Zhu, X. (2025).
\newblock The role of hybrid models in financial decision-making: Forecasting stock prices with advanced algorithms.
\newblock {\em Egyptian Informatics Journal}, 29:100610.

\end{thebibliography}

\end{document}